\documentclass[11pt]{article}

\usepackage{amssymb,amsmath,mathrsfs,epstopdf,slashed,color,tikz}
\usepackage{tcolorbox}
\usetikzlibrary{matrix}
\usetikzlibrary{positioning}
\usepackage{enumerate}
\usepackage{xcolor}

\usepackage{ifpdf}
\ifpdf
\usepackage{graphicx,placeins,hyperref,cite}
\else
\usepackage[dvipdfmx]{graphicx}
\usepackage[dvipdfmx]{hyperref}
\usepackage[numbers]{natbib}
\usepackage{hypernat}
\fi

\allowdisplaybreaks[1]

\makeatletter

\@addtoreset{equation}{section}
\makeatother

\definecolor{greyish}{rgb}{.90,.90,.90}
\definecolor{greyish2}{rgb}{.96,.96,.96}

\begin{document}

\begin{titlepage}

\setcounter{page}{0}

\begin{flushright}
\small
\normalsize
\end{flushright}

\begin{center}

{\Large \bf Quasinormal modes, echoes and the causal structure of the Green's function}

\vskip 1.5cm

{\large Lam Hui${}^{\,a}$\footnote{\href{mailto:lh399@columbia.edu}{\texttt{lh399@columbia.edu}}},
Daniel Kabat${}^{\,b}$\footnote{\href{mailto:daniel.kabat@lehman.cuny.edu}{\texttt{daniel.kabat@lehman.cuny.edu}}}  and
Sam S. C. Wong${}^{\,a,\,c}$\footnote{\href{mailto:scswong@sas.upenn.edu}{\texttt{scswong@sas.upenn.edu}}}}
 
\vskip 0.6cm

${}^a$ Department of Physics, Center for Theoretical Physics, Columbia University, \\
538 West 120th Street, New York, NY 10027, USA \\[5pt]
${}^b$ Department of Physics and Astronomy, Lehman College, City University of New York,\\
250 Bedford Park Blvd.\ West, Bronx NY 10468, USA \\[5pt]
${}^c$ Center for Particle Cosmology, Department of Physics and Astronomy, \\ University of Pennsylvania 209 S. 33rd St., Philadelphia, PA 19104, USA

\vskip 1.5cm
  
\abstract{\normalsize
Quasinormal modes describe the return to equilibrium of a perturbed system, in particular the ringdown phase of a black
hole merger.  But as globally-defined quantities, the quasinormal
spectrum can be highly sensitive to 
global structure, including distant small perturbations to the potential.  In what sense are
quasinormal modes a property of the resulting black hole?  
We explore this question for the linearized perturbation equation
with two potentials having disjoint bounded support.  
We give a composition law for the Wronskian that determines the quasinormal frequencies
of the combined system.  We show that over short time scales the
evolution is governed by the quasinormal frequencies of the individual potentials, while
the sensitivity to global structure can be understood in terms of
echoes.  We introduce an echo expansion of the Green's function and
show that, as expected on general grounds, at any finite time causality limits the number of echoes that can contribute.
We illustrate our results with the soluble example of a pair of
$\delta$-function potentials.  We explicate the causal structure of
the Green's function, demonstrating under what conditions two very different quasinormal
spectra give rise to very similar ringdown waveforms.
}
  
\vspace{1cm}
\begin{flushleft}
\today
\end{flushleft}
 
\end{center}
\end{titlepage}

\setcounter{footnote}{0}
\renewcommand\thefootnote{\mbox{\arabic{footnote}}}

\section{Introduction\label{sect:intro}}
Detection of gravitational waves from binary black hole collisions
\cite{TheLIGOScientific:2016pea} and future precision measurements
provide an unprecedented tool for probing gravity in the strong field
regime.
Tremendous efforts have been made in extracting information
from the emitted gravitational waves. The ringdown phase of a black
hole collision can be described analytically by linearized gravity
around the black hole background \cite{Kokkotas:1999bd,Berti:2009kk}
\footnote{The ringdown discussed in this paper applies
  equally well to mergers of black holes and neutron stars, as long as
  the resulting object is a black hole.}.
The ringing modes of the
merged black hole are known as quasinormal modes (QNMs) 
and their characteristic frequencies and decay are given by a set of 
complex quasinormal frequencies (QNFs). 

Pioneering work by Regge-Wheeler \cite{Regge:1957td} and by Zerilli
\cite{Zerilli:1970se}
established the 
equations describing the (parity odd and even) linearized metric perturbations around a
Schwarzschild black hole. 
In both cases, the relevant equation, after factoring out suitable
angular harmonics, takes the form:
\begin{eqnarray}
(\partial_t^2 - \partial_x^2 + V(x)) \phi (t, x) = 0 \, ,
\end{eqnarray}
where $\phi$ is the perturbation of interest, $x$ represents the (tortoise) radial coordinate and $V(x)$
represents the appropriate potential; the dependence on the angular harmonic number $\ell$ has
been suppressed.
\footnote{The azimuthal harmonic number $m$ is
  typically set to zero. An $m \ne 0$ solution can be
  generated from the $m=0$ solution by a suitable rotation.
}
For perturbations around a Kerr black hole, the relevant equation has
an additional term with a single time derivative (Teukolsky \cite{Teukolsky:1973ha}),
but upon Fourier transforming in time i.e. $\phi \propto e^{-i\omega
  t}$, the resulting equation is not
too different from the above, i.e. $(-\partial_x^2 + \tilde V)\phi =
0$ with the modified $\tilde V$ understood to contain dependence on 
the frequency $\omega$. The excitation and decay of perturbations around Kerr black holes have been studied in \cite{Berti:2006wq,Zhang:2013ksa}.

As an example, the Regge-Wheeler potential takes the form:
$V = (1 - 1/r) [\ell(\ell+1)/r^2 - 3/r^3]$, with the Schwarzschild
radius set to unity, and $r$ and $x$ are related by
$x = r + {\,\rm ln\,} (r - 1)$.
The potential peaks around $|x|$ of order unity, and tapers off as 
$x \rightarrow -\infty$ (where the horizon is located) and as $x
\rightarrow \infty$ (far away from the black hole).
It has long been appreciated that the QNM spectrum is
sensitive to modifications to $V$ \cite{Nollert:1996rf,Nollert:1998ys}, even if such modifications
are localized far away from where the Regge-Wheeler potential
peaks. A concrete example is the potential generated by the
environment around the black hole, such as gas, dark matter and stars \cite{Barausse:2014tra}.
In other words, the QNM spectrum, unlike the bound-state spectra we
encounter in quantum mechanics, is altered significantly even by {\it distant}
modifications to the original potential. Recently, there has been much
attention on distant modifications of this kind, but for exotic objects
such as wormholes, or objects with some sort of a membrane or
firewall just outside the would-be horizon
\cite{Cardoso:2016rao,Cardoso:2016oxy,Ghersi:2019trn,
  Cardoso:2017njb,Wang:2019rcf,
  Price:2017cjr,Mark:2017dnq,Bueno:2017hyj}.
Such objects effectively have two
well-separated potential bumps, one of which is the essentially the
Regge-Wheeler potential.
One interesting outcome of the ringdown computation for these objects
is that despite the vastly different QNM spectrum compared to a normal
black hole, the ringdown waveform of the former is remarkably similar to that of the
latter (e.g. compare Fig. 2 and Fig. 4 of \cite{Cardoso:2016rao}),
except for the presence of echoes over long time scales.
Our goal in this paper is to reconcile these two seemingly
contradictory facts---significantly different QNM spectra, yet rather
similar initial ringdown waveforms. The tool for studying time
evolution is the Green's function. We will explicate the relation between
the QNM spectrum and the Green's function, and introduce an echo
expansion which clarifies the causal structure of the latter. 
We will use the simple example of a potential with two separated delta
functions to illustrate the main ideas.

It should be emphasized that much of the discussion here is not new.
The relation between the QNM spectrum and poles of the Green's
function was discussed for instance in \cite{Leaver:1986gd,Nollert:1992ifk,Andersson:1995zk,Szpak:2004sf}.
An expansion of the Green's function in terms of echoes can be found in
\cite{Mark:2017dnq,Correia:2018apm,Huang:2019veb}.
The use of two disjoint delta functions, or more general disjoint
bumps, as a model potential to illustrate properties of the QNM
can be found in
\cite{Correia:2018apm,Cardoso:2019apo,Li:2019kwa,Buoninfante:2019teo,Huang:2019veb}.
Our goal here is a modest one: to clarify
under what conditions the real time evolution of some localized perturbations 
is sensitive (or insensitive) to the full QNM spectrum. 
Related ideas were discussed by \cite{Mirbabayi:2018mdm}.

An outline of this paper is as follows.  In section \ref{sect:motivation} we present a few simple results for $\delta$-function potentials to motivate the
rest of our analysis.  In section \ref{sect:Greens} we give a general discussion of Green's functions, quasinormal modes and echoes for a pair
of potentials with disjoint bounded support.  In section \ref{sect:delta} we use the soluble example of $\delta$-function potentials to illustrate more
general phenomena.  In section \ref{sect:causal} we study generic
potentials of finite width in more detail and show that the echoes
respect causality.  We conclude in section \ref{sect:conclusions},
where we also discuss under what conditions a small change to the
potential produces a small change to the Green's function.
The appendices contain some supporting material: a review of Green's function solutions to the wave equation (appendix \ref{appendix:Greensolution}),
an alternate derivation of the Green's function for a potential consisting of multiple $\delta$-functions (appendix \ref{appendix:alternate}), and a WKB analysis of the matching
coefficients for a general potential (appendix \ref{appendix:ratio}).

\section{Motivation\label{sect:motivation}}
Quasinormal modes describe the return to equilibrium of a perturbed field.  They
are defined by ``radiative'' boundary conditions which are purely outgoing at
future null infinity (and purely ingoing at any future horizons).  But as globally-defined quantities, quasinormal
modes can be very sensitive to the global structure of a spacetime.  Even small perturbations at large distances can
dramatically change the spectrum of quasinormal modes.

As a simple example, consider
the wave equation in 1+1 dimensions with a $\delta$-function
potential:
\footnote{See discussion in \S \ref{sect:intro} on how an
  equation like this naturally arises in a 3+1 context.}
\begin{equation}
\label{SingleDelta}
\left(\partial_t^2 - \partial_x^2 + V_0 \delta(x)\right) \phi(t,x) = 0
\, .
\end{equation}
Here $V_0$ is a constant which we take to be positive, corresponding
to a repulsive potential, similar to what one encounters in the case
of the black hole.
This wave equation has a unique solution with quasinormal
boundary conditions:
\begin{equation}
\label{SingleDeltaQNM}
\phi(t,x) = {\rm const.} \, e^{-{1 \over 2} V_0 (t - \vert x \vert)}
\, .
\end{equation}
The mode decays with time and we can read off the unique quasinormal
frequency $\omega = - {i \over 2} V_0$ (our frequency convention is $\phi
\propto e^{-i\omega t}$). 
It obeys the appropriate quasinormal boundary conditions at future null infinity, namely that $\phi$ is right-moving (a function
of $t-x$) as $t,x \rightarrow +\infty$ and left-moving (a function of $t+x$) as $t \rightarrow + \infty$,
$x \rightarrow -\infty$.  But the mode grows exponentially at spatial infinity, which suggests that it is very
sensitive to distant perturbations.

\begin{figure}[h]
\center
 \includegraphics[scale=0.65]{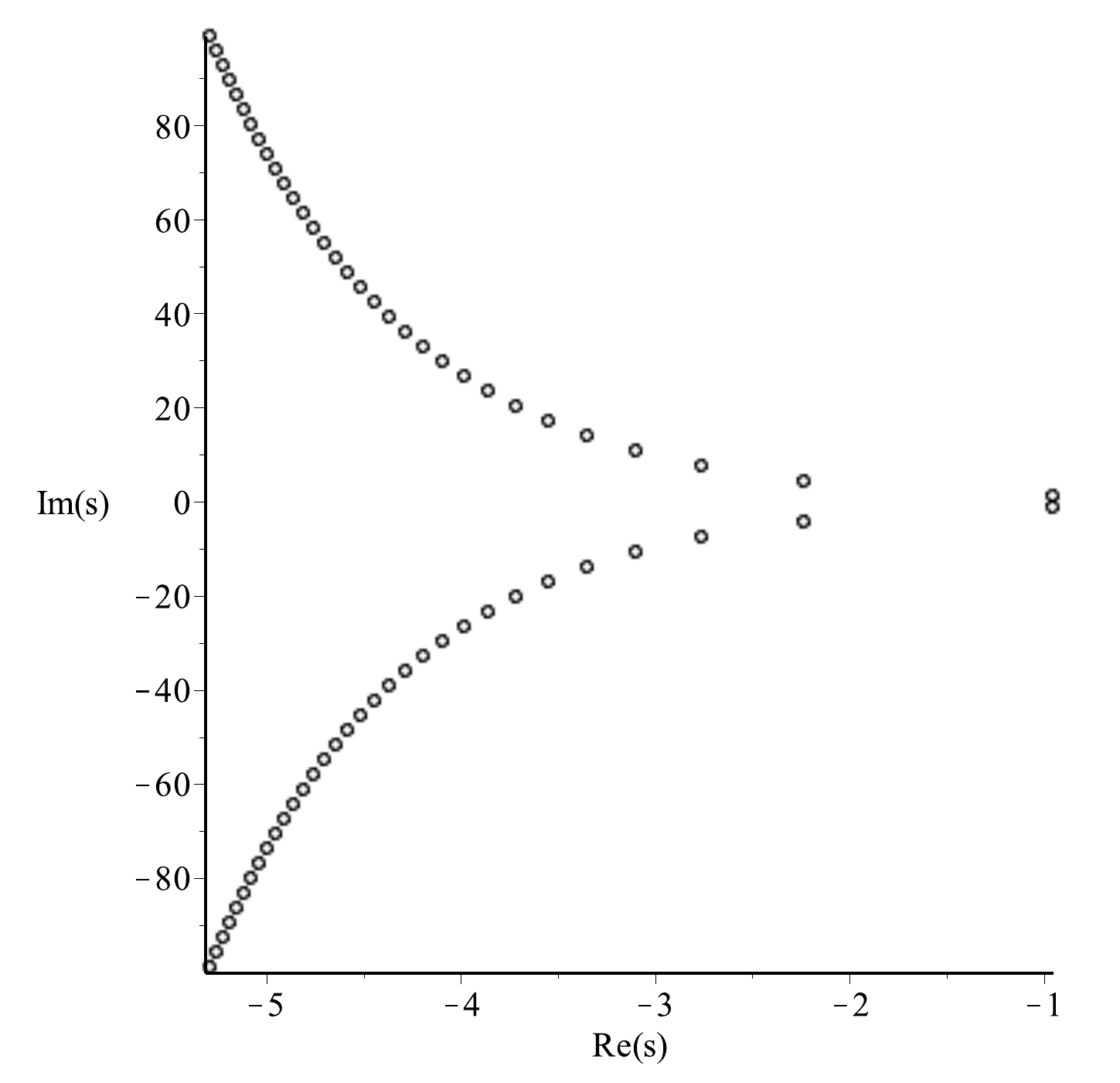}
 \caption{The quasinormal spectrum for a pair of $\delta$-functions with $V_1 = V_2 = L = 1$, obtained by
 setting the Wronskian (\ref{DoubleDeltaW}) to zero.  We show modes with $-100 < {\rm Im} \, s < 100$.\label{fig:zeroes}}
\end{figure}

To see that this is indeed the case, let's compare this to the wave
equation with a pair of $\delta$-function potentials: 
\begin{equation}
\label{DoubleDelta}
\left(\partial_t^2 - \partial_x^2 + V_1 \delta(x) + V_2
  \delta(x-L)\right) \phi(t,x) = 0 \, .
\end{equation}
We will work out the quasinormal spectrum for this problem in section \ref{sect:delta}.  The result is shown in Fig.\ \ref{fig:zeroes}, where we
have plotted the quasinormal frequencies in terms of the Laplace transform variable $s = - i \omega$.  The single mode with
$s = - V_0/2$ splits into the infinite tower of modes shown in the
figure.
Note that the real part of $s$ is always negative for the QNM,
corresponding to a decay with time i.e. $e^{-i\omega t} = e^{st}$. 

\begin{figure}[h]
\center 
\includegraphics[scale=0.65]{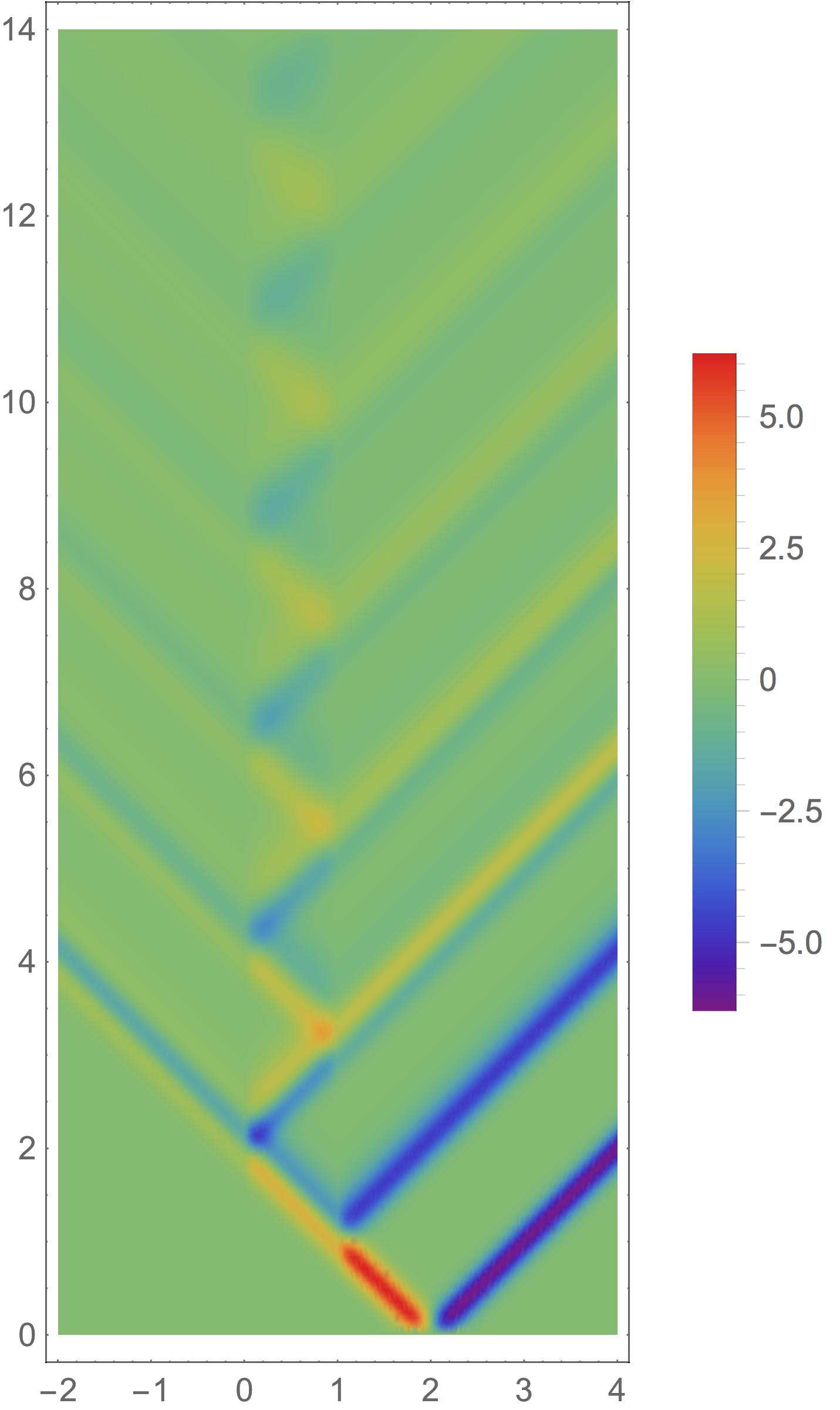}
\caption{A solution to the double $\delta$-function potential,
  obtained using the exact Green's function (\ref{eqn:Gs}), for $V_1 =
  23,\, V_2=11$ and $L=1$. Initial conditions: $\phi(0,x)=0$,
  $\partial_t\phi(0,x)= 500(2-x)e^{-20(x-2)^2}$.
The color coding reflects the value of $\phi$ as indicated on the side bar.
\label{fig:echoes}}
\end{figure}

This dramatic change has a simple interpretation, that it captures the multiple reflections (echoes) which are
present in the two-$\delta$-function system.  These echoes,
illustrated in Fig.\ \ref{fig:echoes}, 
dramatically
change the late-time behavior of the field.

The purpose of this paper is to put these statements on a firm footing, using the language of the retarded Green's function, in the context of two disjoint potentials with bounded support.
The restriction to bounded support is for mathematical convenience, as it will enable us to make sharp statements about the causality properties of the echo expansion in section \ref{sect:causal}.
A discussion of late-time behavior for potentials with more general asymptotic behavior can be found in \cite{Ching:1994bd,Ching:1995tj}.

\section{Green's function and quasinormal modes---general results and
  application to disjoint potentials\label{sect:Greens}}
In this section we begin with a general discussion of the Green's
function 
and quasinormal modes in $1+1$ dimensions,
then specialize to two potentials with disjoint bounded support.

We are interested in the retarded Green's function which satisfies
\begin{eqnarray}
\label{eqn:wave}
&& \left(\partial_t^2 - \partial_x^2 + V(x)\right) G(t,x \vert t',x') = \delta(t-t') \delta(x - x') \\
\label{retardedBC}
&& G = 0 \qquad \hbox{\rm if $t < t'$}
\end{eqnarray}
The Green's function only depends on $t - t'$, so a Laplace transform
\begin{equation}
G(s; x \vert x') = \int_{0^-}^\infty dt \, e^{-st} G(t,x \vert 0,x')
\end{equation}
leads to
\begin{equation}
\label{1D}
\left(-\partial_x^2 + s^2 + V(x)\right) G(s; x \vert x') = \delta(x - x')
\end{equation}
Then we can invert to find
\begin{equation}
\label{invert}
G(t,x \vert t',x') = \int_{c-i\infty}^{c+i\infty} {ds \over 2 \pi i} e^{s(t-t')} G(s; x \vert x')
\end{equation}
(the contour runs parallel to the ${\rm Im} \, s$ axis, but with a positive real part
so that it lies to the right of all singularities).  Our choice of Laplace transform is necessary but not sufficient to give a retarded Green's function, and we still need to ensure that the retarded
boundary condition (\ref{retardedBC}) is satisfied.  Given the inverse transformation (\ref{invert}),
this amounts to demanding that
\footnote{Recall that the integral in (\ref{invert}) runs along a vertical contour to the right of all singularities.  If $G(s; x \vert x')$ is bounded as ${\rm Re} s \rightarrow + \infty$,
then for $t < t'$ we can close the contour to the right at large positive ${\rm Re} \, s$, resulting in a
vanishing $G(t, x|t', x')$. Thus $G(t,x \vert t',x')$ satisfies the boundary condition (\ref{retardedBC}).}
\begin{equation}
\label{1Dbc}
\hbox{\rm $G(s; x \vert x')$ is bounded as ${\rm Re} \, s \rightarrow + \infty$}
\end{equation}

We still have to build a 1-dimensional Green's function satisfying (\ref{1D}), (\ref{1Dbc}).  To solve (\ref{1D}) suppose we find any two linearly-independent solutions
$\phi_\pm(x,s)$ to the homogeneous equation
\begin{equation}
\label{homogeneous}
\left(-\partial_x^2 + s^2 + V(x)\right) \phi(s,x) = 0
\end{equation}
Then a 1D Green's function is \cite{greens}
\begin{equation}
\label{1DG}
G(s; x \vert x') = {1 \over W} \left(\phi_-(s,x) \phi_+(s,x') \theta(x - x')
+ \phi_+(s,x) \phi_-(s,x') \theta(x' - x)\right) \, ,
\end{equation}
where the Wronskian
\begin{equation}
W \equiv W[\phi_-,\phi_+](s) \equiv \phi_- \partial_x \phi_+ -
\phi_+ \partial_x \phi_- 
\end{equation}
is $x$-independent,
and $\theta(x - x')$ is the step function i.e.\ it vanishes if $x - x' < 0$
and equals unity otherwise.\footnote{
The reasoning, briefly, is as follows \cite{greens}. 
Comparing Eq. (\ref{1D}) and Eq. (\ref{homogeneous}), we see that
$G(s; x|x')$ should be a homogeneous solution
if $x > x'$ or $x < x'$. Thus $G(s; x|x') \propto \phi_-(x, s)$ for $x > x'$,
and likewise $G(s; x|x') \propto \phi_+(x, s)$ for $x < x'$. 
The coefficients are dependent on $x'$ and can be figured out
by demanding $G$ produces the correct delta function $\delta (x-x')$
when substituted in the equation of motion.}

The appropriate choice of $\phi_-$ and $\phi_+$ is dictated by (\ref{1Dbc}).  Since we're considering
potentials with bounded support this is simplest to discuss to the left and right of the
potential, where the two independent homogeneous solutions are just growing and decaying exponentials.
To satisfy (\ref{1Dbc}) we take\footnote{We could multiply these solutions by arbitrary non-vanishing functions $f_+(s)$, $f_-(s)$.
Nothing is gained by this generalization since in constructing the Green's function $f_+ f_-$ cancels against the Wronskian.}
\begin{eqnarray}
\label{phip}
&& \phi_+(s,x) = e^{s x} \qquad \,\,\, x < {\rm min} \, {\rm supp} \, V \\
\label{phim}
&& \phi_-(s,x) = e^{-s x} \qquad x > {\rm max} \, {\rm supp} \, V \, ,
\end{eqnarray}
Here the support of the potential is
\begin{equation}
{\rm supp} \, V = \lbrace x \, : \, V(x) \not= 0 \rbrace \, .
\end{equation}

Now suppose the general solution to (\ref{homogeneous}) behaves as
\begin{eqnarray}
\label{general}
&& \phi(s,x) = \left\lbrace
\begin{array}{ll}
\alpha e^{sx} + \beta e^{-sx} & x < {\rm min} \, {\rm supp} \, V \\
\gamma e^{sx} + \delta e^{-sx} & x > {\rm max} \, {\rm supp} \, V
\end{array}\right. \\[5pt]
\label{match}
&& \left(\begin{array}{c} \gamma \\ \delta \end{array}\right) =
\left(\begin{array}{cc} m^{++} & m^{+-} \\ m^{-+} & m^{--} \end{array}\right)
\left(\begin{array}{c} \alpha \\ \beta \end{array}\right)
\end{eqnarray}
(the ``matching coefficients'' $m^{\pm\pm}$ are related to the transmission and reflection coefficients).\footnote{For future reference note that the Wronskian is independent of $x$, which requires
$m^{++} m^{--} - m^{+-}m^{-+} = 1$ so the inverse transformation is
\begin{equation}
\label{inverse}
\left(\begin{array}{c} \alpha \\ \beta \end{array}\right) =
\left(\begin{array}{cc} m^{--} & -m^{+-} \\ -m^{-+} & m^{++} \end{array}\right)
\left(\begin{array}{c} \gamma \\ \delta \end{array}\right)
\end{equation}
}
For instance, $\phi_+$ has $\alpha = 1$ and $\beta = 0$ and thus
$\gamma = m^{++}$ and $\delta = m^{-+}$. 
On the other hand, $\phi_-$ has $\gamma=0$ and $\delta=1$, and
thus $\alpha= -m^{+-}$ and $\beta = m^{++}$. 
Since the Wronskian is independent of $x$ we can compute it for $x > {\rm max} \, {\rm supp} \, V$, where $\phi_+ = m^{++} e^{sx} + m^{-+} e^{-sx}$
and $\phi_- = e^{-sx}$ and hence
\begin{equation}
\label{W}
W = 2sm^{++} \, ,
\end{equation}
where $m^{++}$ is in general a function of $s$.
The Green's function is then
\begin{equation}
\label{generalGreens}
G(t,x \vert t',x') = \int_{c-i\infty}^{c+i\infty} {ds \over 2 \pi i} e^{s(t-t')} {1 \over 2sm^{++}} \left(\phi_-(s,x) \phi_+(s,x') \theta(x - x')
+ \phi_+(s,x) \phi_-(s,x') \theta(x' - x)\right) \, .
\end{equation}
When the Wronskian vanishes the solutions $\phi_+$ and $\phi_-$ are
linearly dependent.  
Recall that $\phi_+$ behaves as $e^{sx}$ to the far left, and
$\phi_-$ behaves as $e^{-sx}$ to the far right; recall also that the
time dependence of the $s$ mode is $e^{st}$. 
To have linearly dependent $\phi_+$ and $\phi_-$ means there is
a single solution which behaves $e^{s(t+x)}$ to the far left and
$e^{s(t-x)}$ to the far right. This is precisely a solution that obeys
the standard quasinormal boundary conditions (i.e. a solution that is
purely outgoing at both boundaries of the domain of interest).
So quasinormal frequencies (in Laplace space) are zeroes of the Wronskian
\cite{Nollert:1992ifk,Ching:1995tj}.

Let us wrap up this general discussion of the Green's function by
noting how it is used to evolve the $\phi$ field. As reviewed in appendix \ref{appendix:Greensolution},
for $t > t'$ we have
\begin{eqnarray}
\label{GreenEvolve}
\phi(t, x) = \int_{-\infty}^{\infty} dx' \left[ G(t, x|t', 
  x') \partial_{t'} \phi(t' , x') - \phi(t', x') \partial_{t'} G(t,
  x|t', x')  \right] \, .
\end{eqnarray}
Thus, given some localized field configuration and its time derivative at $t'$,
the Green's function allows us to evolve the field forward to time 
$t$. It is perhaps not surprising that such a Green's
function knows about the quasinormal spectrum---the influence
from the initial disturbance at $t'$ propagates in an outgoing
manner towards the far left and far right. 

The expressions so far are rather general.  Let us specialize to
the case where $V(x)$
consists of two pieces with disjoint bounded support:
\begin{equation}
V(x) = V_1(x) + V_2(x-L) \, .
\end{equation}
We have in mind that $V_1$ and $V_2$ are separately supported 
around their respective origin i.e. $V_1 (x)$ is non-vanishing around
$x=0$, and $V_2(x-L)$ is non-vanishing around $x-L=0$. 
We have introduced $L > 0$ as an explicit parameter controlling the separation.
It's straightforward to work out a composition law for the Wronskian.  For $x < {\rm min} \, {\rm supp} \, V_1$ we should take the solution
\begin{equation}
\label{left}
\phi_+(s,x) = e^{sx} \, .
\end{equation}
Then for ${\rm max} \, {\rm supp} \, V_1 < x < L + {\rm min} \, {\rm supp} \, V_2$ we will have
\begin{eqnarray}
\label{middle}
\phi_+(s,x) & = & m_1^{++} e^{sx} + m_1^{-+} e^{-sx} \\
\nonumber
& = & e^{sL} m_1^{++} e^{s(x-L)} + e^{-sL} m_1^{-+} e^{-s(x-L)} \, .
\end{eqnarray}
In this form we can use the matching coefficients for $V_2$ to find
the behavior for $x > L + {\rm max} \, {\rm supp} \, V_2$, namely
\footnote{
Because $V_2$ is centered at $x=L$, the analog of Eq. (\ref{general})
for $V_2$ has $x \rightarrow x-L$ on the right hand side, while the
analog of Eq. (\ref{match}) for $V_2$ remains the same.
Using $\alpha = e^{sL} m_1^{++}$ and $\beta = e^{-sL} m_1^{-+}$ allows
one to figure out $\gamma$ and $\delta$. 
}
\begin{equation}
\label{right}
\phi_+(s,x) = \left(m_2^{++} m_1^{++} + e^{-2sL} m_2^{+-}
  m_1^{-+}\right) e^{sx} + \left(m_2^{--} m_1^{-+} + e^{2sL} m_2^{-+}
  m_1^{++}\right)e^{-sx} \, .
\end{equation}
In this range of $x$ we should take $\phi_-(s,x) = e^{-sx}$, so the Wronskian is
\begin{equation}
\label{composition}
W_{1+2} = 2s\left(m_2^{++} m_1^{++} + e^{-2sL} m_2^{+-}
  m_1^{-+}\right) \, .
\end{equation}
Equivalently we have a composition law
\begin{equation}
\label{composition2}
W_{1+2} = {1 \over 2s} W_1 W_2 + 2 s e^{-2sL} m_2^{+-} m_1^{-+} \, .
\end{equation}
Note the exponential dependence on $L$.  The Green's function is
\begin{eqnarray}
\nonumber
G(t,x \vert t',x') & = & \int_{c-i\infty}^{c+i\infty} {ds \over 2 \pi i} e^{s(t-t')} {1 \over 2s\left(m_2^{++} m_1^{++} + e^{-2sL} m_2^{+-} m_1^{-+}\right)} \\[5pt]
\label{compositeGreens}
& & \qquad \left(\phi_-(s,x) \phi_+(s,x') \theta(x - x') + \phi_+(s,x)
    \phi_-(s,x') \theta(x' - x)\right) \, .
\end{eqnarray}

The question is how to interpret (\ref{compositeGreens}).  Compared to (\ref{generalGreens}) the main thing that has changed is the denominator.
We propose that this should be expanded in powers of $e^{-sL}$.
\begin{eqnarray}
\nonumber
G(t,x \vert t',x') & = & \sum_{k=0}^\infty (-1)^k \int_{c-i\infty}^{c+i\infty} {ds \over 2 \pi i} \, {\left(m_2^{+-} m_1^{-+}\right)^k \over 2s\left(m_2^{++} m_1^{++}\right)^{k+1}} e^{s(t-t'-2kL)} \\[5pt]
\label{compositeGreensexpand}
& & \qquad \left(\phi_-(s,x) \phi_+(s,x') \theta(x - x') + \phi_+(s,x) \phi_-(s,x') \theta(x' - x)\right)
\end{eqnarray}
This can be understood as a sum over echoes: if we ignore a possible exponential $s$ dependence of the matching coefficients $m^{\pm\pm}_{1,2}$, the $k^{th}$ term experiences
a time delay of $2kL$.  
Note that for any fixed $t-t'$ the sum over echoes truncates, since for sufficiently large $k$ the contour can be closed to the right and
the integral vanishes.

To make the echoes more explicit we can substitute expressions such as (\ref{left}), (\ref{middle}), (\ref{right}) (and their analogs for $\phi_-$) into
the Green's function.  For example, suppose $x' < {\rm min} \, {\rm supp} \, V_1$ and $x > L + {\rm max} \, {\rm supp} \, V_2$.  Then $\phi_+(x') = e^{sx'}$,
$\phi_-(x) = e^{-sx}$ and
\begin{equation}
G(t,x \vert t',x') = \sum_{k=0}^\infty (-1)^k \int_{c-i\infty}^{c+i\infty} {ds \over 2 \pi i} \, {\left(m_2^{+-} m_1^{-+}\right)^k \over 2s\left(m_2^{++} m_1^{++}\right)^{k+1}} e^{s\big(t-t' - (x - x') - 2kL\big)}
\end{equation}
Suppose the matching coefficients $m^{\pm\pm}_{1,2}$ have no
exponential $s$ dependence; then the $k^{th}$ term in the sum
vanishes if $t - t' < x - x' + 2 k L$. 
This is in accord with the expected time delay for $k$ back-and-forth echoes between sharply-localized potentials separated by a distance
$L$.  For $\delta$-function potentials this conclusion is accurate as
we show in section \ref{sect:delta}.  But for finite-width potentials
there are corrections.  We analyze these corrections in section
\ref{sect:causal} where we show in general that the echoes respect
causality.

\section{$\delta$-function potentials\label{sect:delta}}
In this section we analyze $\delta$-function potentials as a tractable example to illustrate the general phenomenon.

We first treat a single $\delta$-function.  The wave equation is given in (\ref{SingleDelta}), which after a Laplace transform becomes
\begin{equation}
\left(- \partial_x^2 + s^2 + V_0 \delta(x)\right) \phi(s,x) = 0
\end{equation}
This has a pair of solutions, related by $s \rightarrow -s$, given by
\begin{eqnarray}
\label{OneDeltasol}
&& \phi(s,x) = e^{sx} + {V_0 \over s} \sinh (sx) \theta(x) \\
&& \phi(s,x) = e^{-sx} + {V_0 \over s} \sinh(sx) \theta(x)
\end{eqnarray}
Comparing to (\ref{general}) we read off the matching coefficients
\begin{equation}
\begin{array}{ll}
m^{++} = 1 + {V_0 \over 2s} & \quad m^{-+} = - {V_0 \over 2s} \\
m^{+-} = {V_0 \over 2s} & \quad m^{--} = 1 - {V_0 \over 2s}
\end{array}
\end{equation}
From (\ref{W}) the Wronskian is
\begin{equation}
\label{SingleDeltaW}
W = 2sm^{++} = 2s+V_0
\end{equation}
and the Green's function is
\begin{equation}
\label{SingleDeltaG}
G(t,x \vert t',x') = \int_{c-i\infty}^{c+i\infty} {ds \over 2 \pi i} \, {1 \over 2s + V_0} e^{s(t - t')} \left(\phi_-(s,x) \phi_+(s,x') \theta(x - x')
+ \phi_+(s,x) \phi_-(s,x') \theta(x' - x)\right)
\end{equation}
Here solutions $\phi_\pm$ with the behavior (\ref{phip}), (\ref{phim}) are given by\footnote{The first line is the same as (\ref{OneDeltasol}).
The second can be obtained by $x \rightarrow -x$ or by using (\ref{inverse}).}
\begin{eqnarray}
&& \phi_+(s,x) = e^{sx} + {V_0 \over s} \sinh(sx) \theta(x) \\
&& \phi_-(s,x) = e^{-sx} - {V_0 \over s} \sinh(sx) \theta(-x)
\end{eqnarray}
As expected the Wronskian (\ref{SingleDeltaW}) has a zero at the quasinormal frequency $s = - V_0 / 2$.  The contour integral (\ref{SingleDeltaG}) gives
\begin{equation}
\label{SingleDeltaG2}
G(t,x \vert t',x') = \left\lbrace
\begin{array}{ll}
0 & \quad t - t' < \vert x - x' \vert \\
{1 \over 2} & \quad \vert x - x' \vert < t - t' < \vert x \vert + \vert x' \vert \\
{1 \over 2} \exp \left(-{1 \over 2} V_0 (t - t' - \vert x \vert - \vert x' \vert)\right) & \quad t - t' > \vert x \vert + \vert x' \vert
\end{array}\right.
\end{equation}
or alternatively
\begin{equation}
G(t,x \vert t',x') = {1 \over 2} \theta(t - t' - \vert x - x' \vert) + {1 \over 2} \left(e^{-{1 \over 2} V_0 (t - t' - \vert x \vert - \vert x' \vert)} - 1\right) \theta(t - t' - \vert x \vert - \vert x' \vert)
\end{equation}
The first line of (\ref{SingleDeltaG2}) shows that the Green's function vanishes unless $(t,x)$ lies in the causal future of $(t',x')$.\footnote{That is, it vanishes unless
$(t,x) \in {\cal J}^+(t',x')$. 
}  The second line recovers the Green's function without a potential, in the case where causal curves from $(t',x')$ to $(t,x)$ do not touch $x = 0$.\footnote{That is, when the causal
diamond ${\cal J}^+(t',x') \cap {\cal J}^-(t,x)$ does not touch
$x=0$.
Note that if $x$ and $x'$ are on
opposite sides of the delta function at the origin, then
$|x - x'| = |x| + |x'|$ and the possibility in the second line is never realized, i.e. either the Green's function vanishes, or
else the delta function leaves an
imprint on it.
}  In the last line -- the regime where scattering off the
potential is permitted by causality -- the Green's function depends on
$V_0$.

To extend this to the pair of $\delta$-functions (\ref{DoubleDelta}) we make use of (\ref{composition}), which tells us that the Wronskian for the combined system is (recall $L > 0$)
\begin{equation}
\label{DoubleDeltaW}
W = {1 \over 2s}\left[\left(2s + V_1\right)\left(2s + V_2\right) - V_1 V_2 e^{-2sL}\right]
\end{equation}
Zeroes of the Wronskian determine quasinormal frequencies (the dictionary is $\omega = i s$).  The individual potentials had zeroes at $s = -V_1/2$ and $s = -V_2/2$, but
the combined system has an infinite number of quasinormal frequencies.
This is illustrated in Fig.\ \ref{fig:zeroes} for the case $V_1 = V_2
= L = 1$.

The Green's function for the combined system is given by (\ref{compositeGreens}),
\begin{eqnarray}
\nonumber
G(t,x \vert t',x') & = & \int_{c-i\infty}^{c+i\infty} {ds \over 2 \pi i} e^{s(t-t')} {2s \over \left(2s + V_1\right)\left(2s + V_2\right) - V_1 V_2 e^{-2sL}} \\[5pt]
\label{TwoDeltacompositeGreens}
& & \qquad \left(\phi_-(s,x) \phi_+(s,x') \theta(x - x') + \phi_+(s,x)
    \phi_-(s,x') \theta(x' - x)\right) \, ,
\end{eqnarray}
where solutions $\phi_\pm$ satisfying (\ref{phip}), (\ref{phim}) are\footnote{The first line is a special case of (\ref{left}), (\ref{middle}), (\ref{right}).  The second follows from
(\ref{inverse}).} 
\begin{eqnarray}
&& \phi_+(s,x) = e^{sx} + {V_1 \over s} \sinh(sx) \theta(x) + {V_2 \over s} \Big(e^{sL} + {V_1 \over s} \sinh(sL)\Big)\sinh\big(s(x-L)\big) \theta(x-L) \\
\nonumber
&& \phi_-(s,x) = e^{-sx} - {V_2 \over s} e^{-sL} \sinh\big(s(x-L)\big)
   \theta(L-x) - {V_1 \over s} \Big(1 + {V_2 \over s} e^{-sL}
   \sinh(sL)\Big)\sinh(sx) \theta(-x) \, .
\end{eqnarray}
This leads to
\begin{align} \label{eqn:Gs}
 G(t,x\vert t',x') &= \int_{c-i\infty}^{c+i\infty} {ds \over 2 \pi i} e^{s(t-t')} \Bigg[\frac{1}{2s}e^{-s|x-x'|}  \nonumber \\
     & \quad + \frac{ V_1(2s+V_2)e^{-s (|x|+ |x'|)} - V_1V_2 e^{-s(|x| + L+|L-x'|)} }{2s\left[V_1 V_2 e^{-2sL}- (2s +V_1 )(2s+V_2)  \right]}  \nonumber \\
       & \quad + \frac{ V_2(2s+V_1)e^{-s (|L-x|+ |L-x'|)} - V_1V_2
         e^{-s(|L-x| + L+|x'|)} }{2s\left[V_1 V_2 e^{-2sL}- (2s +V_1
         )(2s+V_2)  \right]}\Bigg] \, .
\end{align}
An alternate derivation of this result can be found in appendix \ref{appendix:alternate}.

As we now show, this can be interpreted in terms of echoes by expanding in powers of $e^{-sL}$.
The inverse Laplace transform in the first line just gives the usual Green's function for the operator $\partial_t^2 - \partial_x^2$.
\begin{equation}
{\cal L}^{-1}\left[  \frac{1}{2s}e^{-s|x-x'|}   \right] = \frac{1}{2}\theta(t-t'-|x-x'|)
\end{equation}
The second and third lines are sensitive to the potential and have poles at the quasinormal frequencies.
Let's focus on the first term in the second line,
\begin{align}
 & \frac{ V_1(2s+V_2)e^{-s (|x|+ |x'|)}  }{2s\left[V_1 V_2 e^{-2sL}- (2s +V_1 )(2s+V_2)  \right]}  \nonumber \\
 \label{TwoDeltaExpand}
 =\;&- \sum_{k=0}^{\infty}  \frac{e^{-s(2kL+|x|+|x'|)}}{2s \left(s+\frac{V_2}{2}\right)^k \left(s+\frac{V_1}{2}\right)^{k+1}} \left(\frac{V_2}{2}\right)^k  \left(\frac{V_1}{2}\right)^{k+1}
\end{align}
where we have expanded in $e^{-2sL}$.  Now the poles are of finite order and can be inverse Laplace transformed. 
\begin{align}
{\cal L}^{-1}[ \cdots ] = \;& \frac{1}{2}\left( e^{-\frac{V_1}{2}(t-t'-|x|-|x'|) } -1\right)\theta(t-t'-|x|-|x'|) \nonumber \\
  &-\sum_{k=1}^{\infty} \frac{1}{2}\Bigg[  \frac{1}{k!} \frac{d^k}{ds^k} \left.\left(  \frac{e^{s(t-t'-2kL-|x|-|x'|)}}{s \left(s+\frac{V_2}{2}\right)^k }\right) \right|_{s=-\frac{V_1}{2}} \left(\frac{V_2}{2}\right)^k  \left(\frac{V_1}{2}\right)^{k+1} \nonumber \\
  & \qquad + \frac{1}{(k-1)!} \frac{d^{k-1}}{ds^{k-1}} \left.\left(  \frac{e^{s(t-t'-2kL-|x|-|x'|)}}{s \left(s+\frac{V_1}{2}\right)^{k+1} }\right) \right|_{s=-\frac{V_2}{2}} \left(\frac{V_2}{2}\right)^k  \left(\frac{V_1}{2}\right)^{k+1} \nonumber \\
  &\qquad  +1 \Bigg] \theta(t-t'-2kL-|x|-|x'|).
\end{align}
The inverse Laplace transform of the other terms in (\ref{eqn:Gs}) can be handled in a similar manner and the resulting Green's function is 
\begin{align}
G(t,x\vert t',x') = \; & \frac{1}{2}\theta(t-t'-|x-x'|)  \nonumber \\
  &+ \Bigg\{  \frac{1}{2}\left( e^{-\frac{V_1}{2}(t-t'-|x|-|x'|) } -1\right)\theta(t-t'-|x|-|x'|) \nonumber \\
  &-\sum_{k=1}^{\infty} \frac{1}{2}\Bigg[  \frac{1}{k!} \frac{d^k}{ds^k} \left.\left(  \frac{e^{s(t-t'-2kL-|x|-|x'|)}}{s \left(s+\frac{V_2}{2}\right)^k }\right) \right|_{s=-\frac{V_1}{2}} \left(\frac{V_2}{2}\right)^k  \left(\frac{V_1}{2}\right)^{k+1} \nonumber \\
  & \qquad + \frac{1}{(k-1)!} \frac{d^{k-1}}{ds^{k-1}} \left.\left(  \frac{e^{s(t-t'-2kL-|x|-|x'|)}}{s \left(s+\frac{V_1}{2}\right)^{k+1} }\right) \right|_{s=-\frac{V_2}{2}} \left(\frac{V_2}{2}\right)^k  \left(\frac{V_1}{2}\right)^{k+1} \nonumber \\
  &\qquad  +1 \Bigg] \theta(t-t'-2kL-|x|-|x'|) \nonumber \\
   &+\sum_{k=0}^{\infty}\frac{1}{2} \Bigg[ \frac{1}{k!} \frac{d^k}{ds^k} \left.\left(  \frac{e^{s(t-t'-(2k +1)L-|x|-|x'-L|)}}{s \left(s+\frac{V_2}{2}\right)^{k+1} }\right) \right|_{s=-\frac{V_1}{2}} \left(\frac{V_1V_2}{4}\right)^{k+1}  \nonumber \\
   & \quad + \frac{1}{k!} \frac{d^k}{ds^k} \left.\left(  \frac{e^{s(t-t'-(2k+1)L-|x|-|x'-L|)}}{s \left(s+\frac{V_1}{2}\right)^{k+1} }\right) \right|_{s=-\frac{V_2}{2}} \left(\frac{V_1V_2}{4}\right)^{k+1} +1 \Bigg]  \nonumber \\
  & \quad   \times \theta(t-t'-(2k+1)L-|x|-|x'-L| )  \nonumber \\ \label{eqn:Gdelta}
  & + \left(\begin{array}{c} 
  V_1 \leftrightarrow V_2 \\
  |x| \leftrightarrow |x-L|\\
    |x'| \leftrightarrow |x'-L|\\
  \end{array}\right)\Bigg\}  
\end{align}
We will refer to this
rewriting of the Green's function (Eq. (\ref{eqn:Gs})) in the form of
Eq. (\ref{eqn:Gdelta}) as an {\it echo expansion}.

\begin{figure}[h]
\center
\includegraphics[scale=0.5]{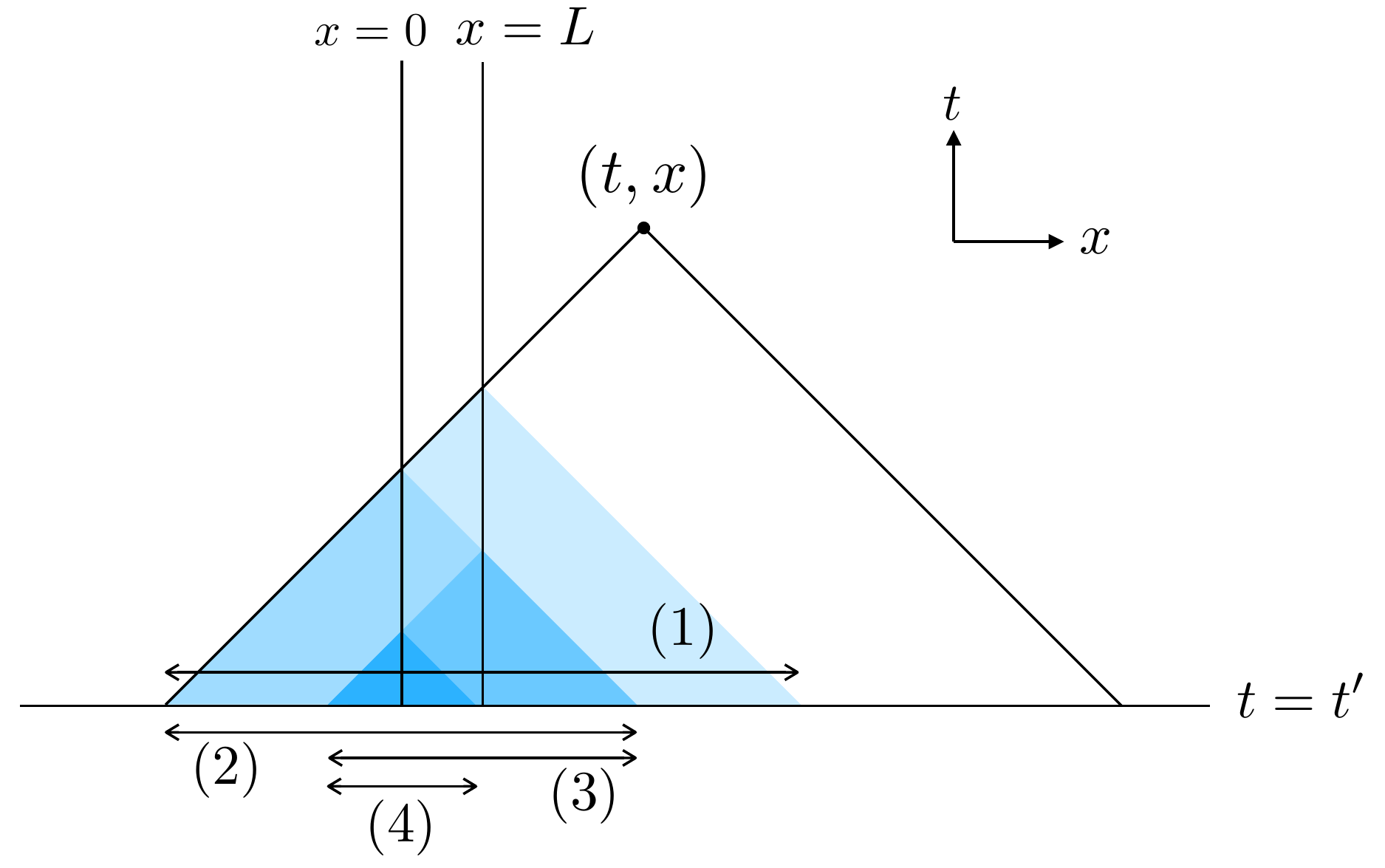}
\caption{
Domain of dependence of the Green's function at $(t,x)$.  The
intervals along the horizontal axis correspond to the following step functions 
(i.e.\ the intersection of the shaded regions with the time$=t'$ surface indicates
the range of $x'$ that gives a non-zero contribution, for $t,x$
fixed as shown):
\newline \hspace*{1cm} (1) \quad $\theta(t-t'-|x-L|-|x'-L|)$, \newline \hspace*{1cm} (2) \quad $\theta(t-t'-|x|-|x'|), \; \theta(t-t'-L-|x-L|-|x'|)$, \newline \hspace*{1cm} (3) \quad $\theta(t-t'-2L-|x-L|-|x'-L|), \; \theta(t-t'-L-|x|-|x'-L|)$, \newline \hspace*{1cm} (4) \quad $\theta(t-t'-2L-|x|-|x'|), \; \theta(t-t'-3L-|x-L|-|x'|)$ \newline These
step functions have the interpretation of sequential scattering off \newline \hspace*{1cm} (1) \quad just $V_2$, \newline \hspace*{1cm} (2) \quad just $V_1$, \; $V_1$ $\rightarrow$ $V_2$, \newline \hspace*{1cm} (3) \quad $V_2$ $\rightarrow$ $V_1$ $\rightarrow$ $V_2$, \; $V_2$ $\rightarrow$ $V_1$, \newline \hspace*{1cm} (4) \quad $V_1$ $\rightarrow$ $V_2$ $\rightarrow$ $V_1$, \; $V_1$ $\rightarrow$ $V_2$ $\rightarrow$ $V_1$ $\rightarrow$ $V_2$\label{fig:1}}
\end{figure}

\begin{figure}[h]
\center
\includegraphics[scale=0.7]{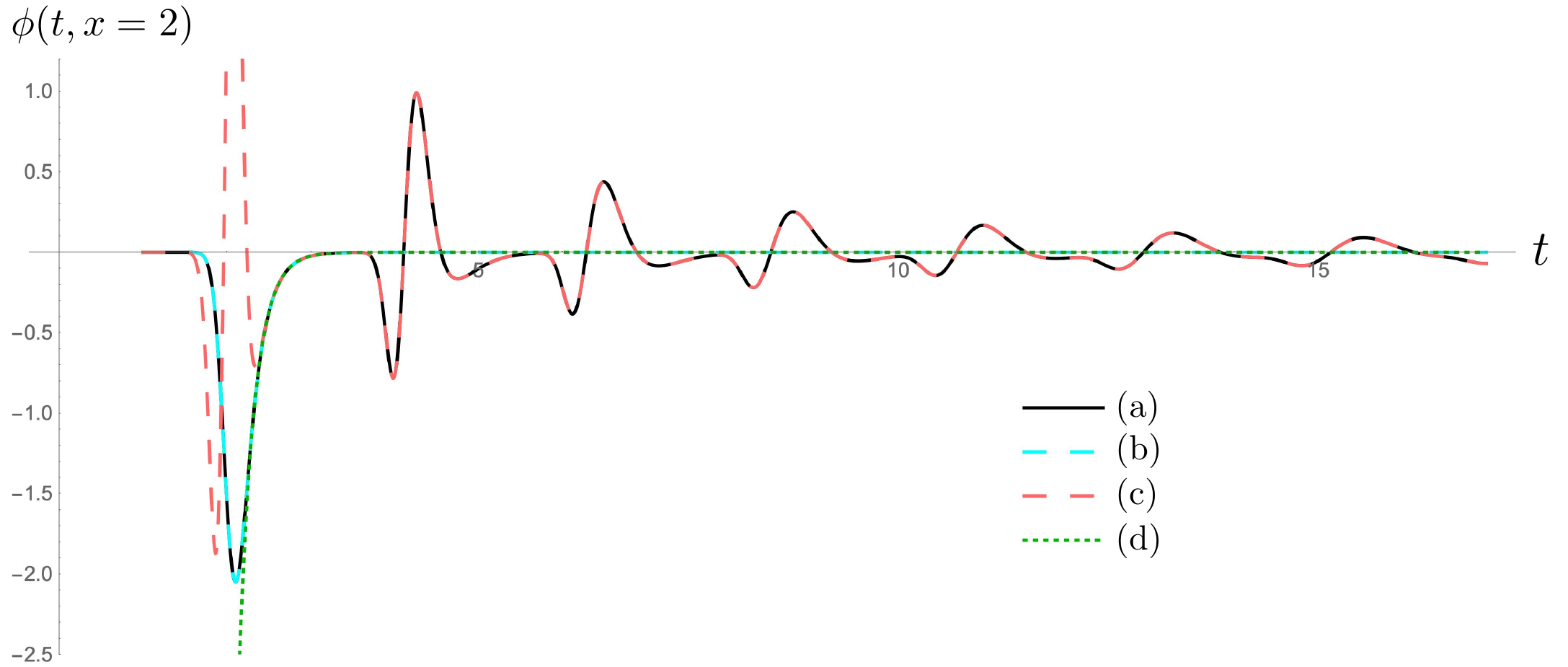}
\caption{(a) the waveform of the solution in Figure \ref{fig:echoes} at $x=2$. (b) a solution obtained using the same Cauchy data but the Green's function of $V_2$ only; note that the resulting waveform gets the first bump right but vanishes thereafter. (c) a fitting of the full waveform using the first 30 quasinormal modes (the modes with smallest $\vert {\rm Re} \, s \vert$) of the combined system. (d) the quasinormal mode of $V_2$ alone, fitted to the first reflected wave.\label{fig:waveform}} 
\end{figure}

This Green's function may look unwieldy but can be interpreted as follows.
First note that the expression in $\{\cdots\}$ encodes the quasinormal frequencies of the {\em individual} potentials.
This is easiest to see in Laplace space, where for example each term in (\ref{TwoDeltaExpand}) only has (higher-order) poles at the individual quasinormal frequencies 
$-{V_1}/{2},\; -{V_2}/{2}$.  The $k^{th}$ term in the sum comes from
$k$ back-and-forth bounces between the potentials, as can be seen from the factors $e^{-s2kL}$ in Laplace space which produce a time delay $2kL$
in real space.  These time delays appear in the step functions which encode the causality properties of the echo expansion.
This is illustrated in Fig.\ \ref{fig:1} which shows the domain of dependence of the field at a particular point $(t,x)$ and indicates how initial data in various
regions at time $t'$ can propagate by multiple scattering to influence the field at $(t,x)$.  Note that there are regions of $(t,x)$ where, by causality, only one of the potentials
can contribute to the Green's function.  In these regions the field evolves in time as though there was only a single potential, even though the quasinormal spectrum of the combined
system is very different from that of a single $\delta$-function.  This is illustrated in Fig.\ \ref{fig:waveform}, where we show (a) the full waveform, (b) the same Cauchy
data but evolved with the Green's function for a single potential.

The last feature we want to mention in this section is the meaning of the quasinormal frequencies of the combined system.
A collection of $\delta$-function potentials provides a particularly simple example since the Green's function only has poles (not branch cuts) in Laplace space.
This means that at sufficiently late times -- once causal curves from the Cauchy data can see the entire potential -- the field may be expanded as a linear combination of quasinormal modes.\footnote{The argument runs as follows.  Eqn.\ (\ref{TwoDeltacompositeGreens}) has poles at the quasinormal frequencies of the combined system.  We can make this explicit by writing
\begin{equation}
\label{PartialFraction}
{1 \over W} = {2s \over (2s + V_1)(2s + V_2) - V_1V_2 e^{-2sL}} = \sum_i \frac{a_i}{s-s_i}
\end{equation}
Once $t$ is large enough that we can close the contour in (\ref{TwoDeltacompositeGreens}) to the left, the Green's function is a linear combination of
quasinormal modes.  By examining the exponentials in the numerator of (\ref{eqn:Gs}), we see that this happens once causal curves from $(t',x')$ to $(t,x)$ can see the
entire potential.  So by (\ref{GreenEvolve}), once causal curves from the Cauchy data can see the entire potential the field is a linear combination of quasinormal modes.
Higher-order poles in (\ref{PartialFraction}) would give derivatives of quasinormal modes with however the same exponential fall-off. A more detailed discussion of QNM expansion of waveforms can be found in \cite{Ching:1995rt,Ching:1999fb}.}
This can be seen in Fig.\ \ref{fig:waveform}, where (c) shows a fit of the full waveform to a linear combination of
the first 30 quasinormal modes of the combined system (the 30 modes with smallest $\vert {\rm Re} s \vert$).\footnote{The Cauchy data in Figure \ref{fig:echoes} extends out
to about $x = 2.5$.  From the arguments in the previous footnote the expansion in quasinormal modes is justified after $t \approx 4.5$, when all terms in (\ref{TwoDeltacompositeGreens}) can be closed to the left.  But the term $\sim e^{-s(|L-x| + L+|x'|)}$ in (\ref{TwoDeltacompositeGreens}), which can be closed to the left once causal curves touch $V_2$, makes the
dominant contribution to Fig.\ \ref{fig:waveform} while all other terms are exponentially suppressed.  So in practice the fit to quasinormal modes is very good after $t \approx 2.5$.}  At late times the lowest
QNF dominates.  But at intermediate times multiple QNFs contribute, in a way that depends on the initial conditions for the field.

But even at late times the quasinormal frequencies
of the individual potentials play a role.  As discussed above each term in the echo expansion only has (higher-order) poles at the quasinormal
frequencies of the individual potentials.  Each echo therefore decays according to a linear combination of the quasinormal frequencies of the
individual potentials, as can be seen explicitly in the coefficients of the step functions in (\ref{eqn:Gdelta}).
As a somewhat trivial example, this is illustrated
for the first reflected wave in Fig.\ \ref{fig:waveform}, where (d) shows a fit to the quasinormal mode of $V_2$ alone.

\section{Echoes and causality} \label{sect:causal}
In the last section we saw that in the case of a two-delta-function
potential, each echo contribution (labeled by $k$) to the Green's
function comes with a corresponding step function. The step
function enforces causality, in the sense that in order for
the step function to not vanish, $t-t'$ (the
separation between the time of interest $t$ and the initial time $t'$)
must be smaller than the {\it expanded} spatial distance between the point of
interest $x$ and the ``source'' $x'$---{\it expanded} by the extra
distance traversed by $k$ echoes.
We wish show in this section that essentially the same statements
apply to more general disjoint potentials. As a bonus, we will see
how the individual potentials govern time evolution over sufficiently
short time scales.

We consider potentials of the form $V(x) = V_1(x) + V_2(x)$ where $V_1$ and $V_2$ have disjoint bounded support.  We take ${\rm supp} \, V_1 = (x_1,x_2)$ and ${\rm supp} \, V_2 = (x_3,x_4)$
as shown below.
\\
\centerline{\includegraphics[scale=0.85]{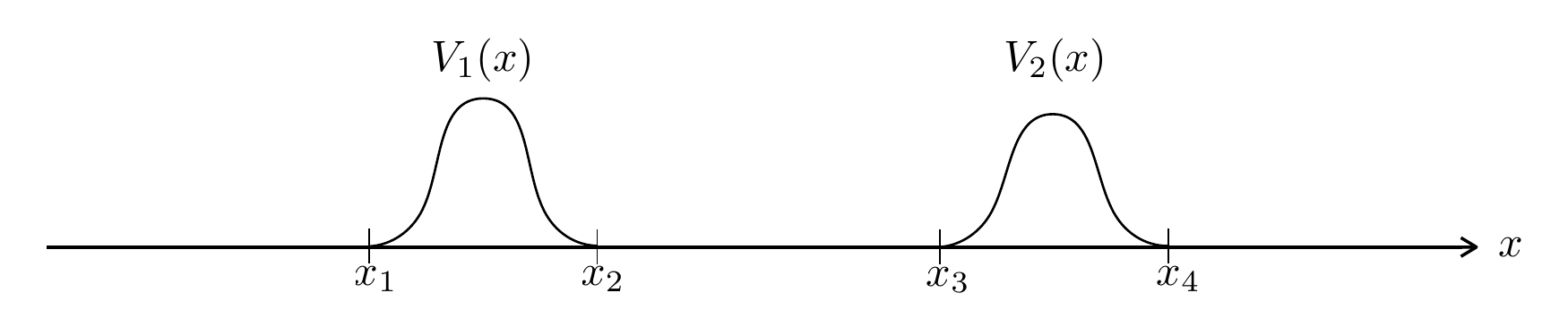}}
The separation between potentials is given by $x_3 - x_2$, so the parameter $L$ has no role to play and will be dropped from all formulas in this section.
Instead of an expansion in powers of $e^{-2sL}$ we perform an expansion in powers of the small quantity
\begin{equation}
{m_1^{-+} m_2^{+-} \over m_1^{++} m_2^{++}} \,\sim\, e^{- 2 s (x_3 - x_2)} \qquad \hbox{\rm at large ${\rm Re} \, s$}
\end{equation}
(see equation (\ref{eqn:ratioasymp}) below).

\begin{figure}[h]
\center
\includegraphics[scale=0.55]{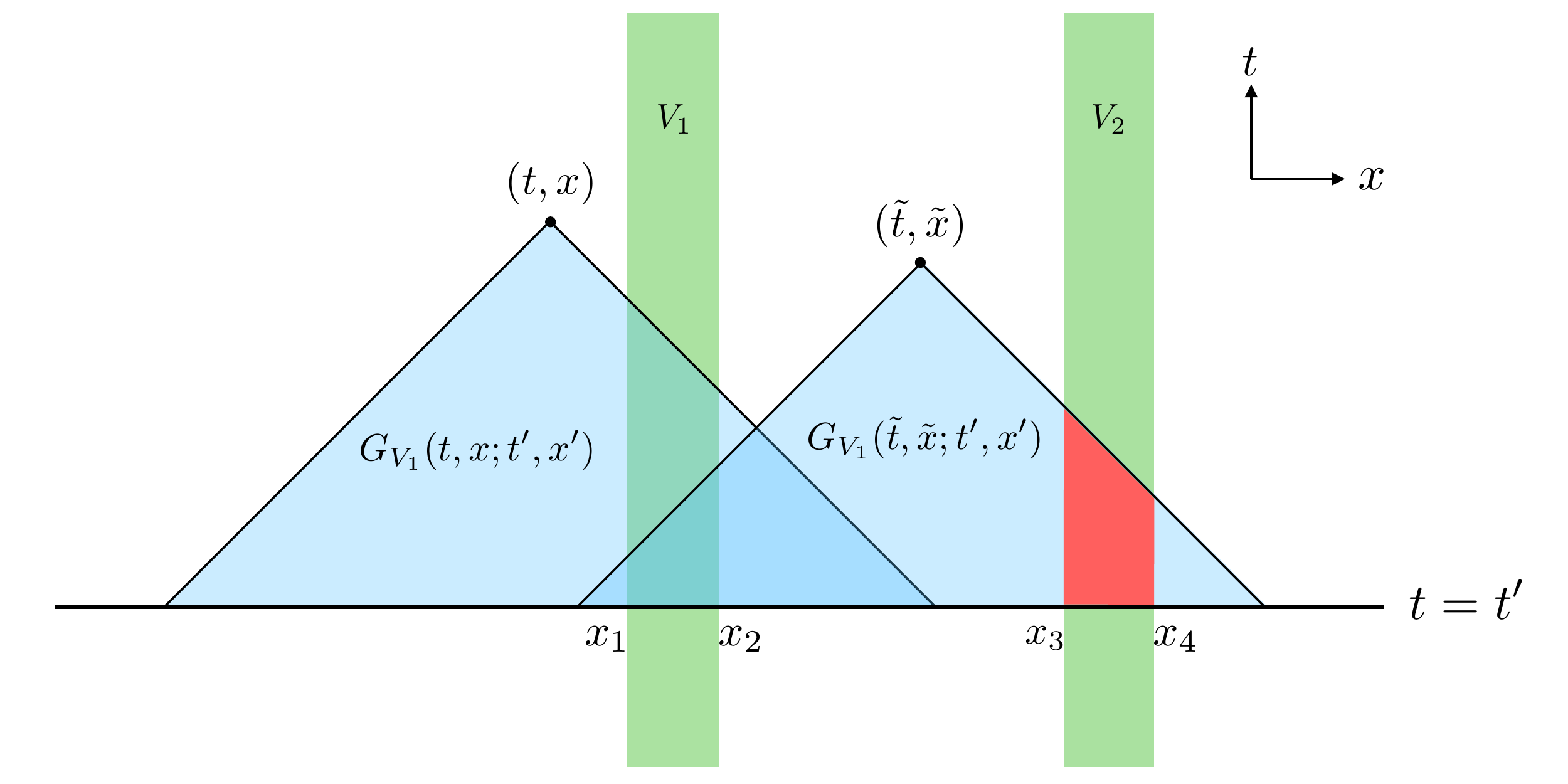}
\caption{Cauchy data is given on the time slice $t'$. The retarded
  Green's function for $V_1$, $G_{V_1}$, has support in the past
  lightcone of the indicated points. The red area is where $G_{V_1}$
  {\it fails to solve} $(\partial_t^2-\partial_x^2 +V_2(x)) G_{V_1} =
  \delta(t-t')\delta(x-x')$.
\label{fig:2}}
\end{figure}

The Wronskian for the combined system $V_1 + V_2$ is given by (\ref{composition}) and (\ref{composition2}),
\begin{eqnarray}
W_{1+2} & = & 2s\left(m_2^{++} m_1^{++} + m_2^{+-} m_1^{-+}\right) \\
\nonumber
& = & {1 \over 2s} W_1 W_2 + 2 s m_2^{+-} m_1^{-+}
\end{eqnarray}
So we know how the Wronskian for the individual potentials is related to the Wronskian for the sum.  From (\ref{compositeGreens}) the Green's function for the combined system is
\begin{equation}
\label{eqn:GW}
G(t,x \vert t',x') = \int_{c-i\infty}^{c+i\infty} {ds \over 2 \pi i} e^{s(t-t')} {1 \over W_{1+2}} \left(\phi_-(s,x) \phi_+(s,x') \theta(x - x') + \phi_+(s,x) \phi_-(s,x') \theta(x' - x)\right)
\end{equation}
We claim that: 
\begin{tcolorbox}[colframe=white,arc=0pt,colback=greyish2]
For any finite period of time, one should treat the Wronskian in the Green's function as
\begin{equation} \label{eqn:Wexpan}
 \frac{1}{W_{1+2}} = \frac{1}{2s m_2^{++}m_1^{++}} \sum_{k=0}^{\infty} (-1)^k \left(\frac{m_2^{+-}m_1^{-+}}{m_2^{++}m_1^{++}}\right)^k = \frac{2s}{ W_1 W_2} \sum_{k=0}^{\infty} (-1)^k \left(\frac{4s^2 m_2^{+-}m_1^{-+}}{W_1 W_2}\right)^k
\end{equation}
\begin{enumerate}[I.]
\item
The sum over $k$ can be understood as a sum over the number of back-and-forth bounces between the two potentials.
Causality is encoded in the sum, and at any finite time the sum truncates to the maximum number of bounces permitted by causality.\label{claim1}
\item
Over sufficiently short times causality forbids a back-and-forth bounce, and one would only observe the quasinormal frequencies associated with the individual potentials $V_1$ and $V_2$.\label{claim2}
\item
Suppose that on the Cauchy surface (the time=$t'$ surface)
the past lightcone of $(x,t)$ only includes one of the potentials, say
$V_1$.  (More precisely, what matters is that on the Cauchy surface the past lightcone excludes the support of $V_2$.)  Then the full Green's function at $(x,t)$ agrees with the Green's function just for $V_1$.\label{claim3}
\end{enumerate}
\end{tcolorbox}
All these properties can be seen in the explicit Green's function for
the case with two delta-function potentials (\ref{eqn:Gdelta}).  
Let us continue the analysis for general potentials with disjoint
bounded support to verify these properties.  The basic fact we'll need is
that at large positive ${\rm Re} \, s$ the matching coefficients have the behavior
\begin{equation}
\label{eqn:ratioasymp}
{m_1^{-+} \over m_1^{++}} \,\sim\, e^{2 s x_2} \hspace{2cm}
{m_2^{+-} \over m_2^{++}} \,\sim\, e^{-2 s x_3}
\end{equation}
where we are only retaining the leading exponential dependence on $s$.  We establish this behavior by a WKB analysis in appendix \ref{appendix:ratio}.  We now verify each claim in turn.
\begin{enumerate}[I.]
\item
With the expansion (\ref{eqn:Wexpan}), the $k=0$ term in the Green's function (\ref{eqn:GW}) is only non-zero when $t-t' \ge f(x,x')$ for some $f(x,x')$ when the contour starts to close to the left. Terms with higher $k$
come with a factor
\begin{equation}
\left(\frac{m_2^{+-}m_1^{-+}}{m_2^{++}m_1^{++}} \right)^k \overset{s \to \infty}{\sim}  e^{-2ks (x_3-x_2)},
\end{equation}
So the $k^{th}$ term vanishes (the contour can be closed to the right) for $t-t'<f(x,x')+2k(x_3-x_2)$.  Since $2k(x_3-x_2)$ is exactly the minimal time for waves to make $k$ round trips between the two potentials, we can interpret $k$ as the number of back-and-forth bounces.  Moreover causality is preserved and the sum is truncated at the maximum number of bounces permitted by causality.
\item
Claim \ref{claim2} is an easy corollary.  For sufficiently short times ($t-t' < f(x,x') + 2(x_3-x_2)$) only the $k=0$ term in the Green's function contributes, so we can replace
\begin{equation}
{1 \over W_{1+2}} \quad {\rm with} \quad {2 s \over W_1 W_2}
\end{equation}
After making this replacement poles only arise from the zeroes of $W_1$ and $W_2$, i.e.\ the quasinormal frequencies of the individual potentials.
\item
To verify claim \ref{claim3} we employ the uniqueness theorem.  Suppose we provide Cauchy data $\phi(t',x')$, $\partial_{t'} \phi(t',x')$ on a time slice $t'$.
Consider a point $(x,t)$ with $t>t'$ satisfying $t-t'<x_3-x$.  Then the past lightcone of $(x,t)$ doesn't make contact with $V_2(x)$ to the future of the Cauchy surface.
This is illustrated in Fig.\ \ref{fig:2}. One can use the retarded Green's function for $V_1$ to construct a field by
\begin{equation}
\phi(t,x) = \int dx' \left( G_{V_1}(t,x\vert t',x') \partial_{t'}\phi(t',x') -\partial_tG_{V_1}(t,x\vert t',x') \phi(t',x')\right).
\end{equation}
It is obvious that this solves the equation of motion for $t-t'<x_3-x$
and also satisfies the initial conditions at $t'$. Due to the
uniqueness theorem this solution is unique therefore the retarded
Green's function $G_{V_1}$ we used must equal the full retarded
Green's function $G_{V_1+V_2}$ within this spacetime domain. Mismatch
between  $G_{V_1}$ and $G_{V_1+V_2}$ occurs when $t-t'>x_3-x$ since
$G_{V_1}$  does not solve the Green's function equation of motion in
the red region depicted in Fig.\ \ref{fig:2}.
\end{enumerate}

\section{Discussion and conclusion\label{sect:conclusions}}
In this work we have explored simple potential models to develop a better understanding of quasinormal modes.  There are several lessons we can draw from our analysis.

First, the basic object of interest is the retarded Green's function.  In real space the retarded Green's function vanishes for $t < t'$, which in Laplace space requires that we impose the
boundary conditions (\ref{phip}), (\ref{phim}).  In this approach to constructing the Green's function, the quasinormal frequencies arise as derived quantities: they are simply the poles of the Green's function
in Laplace space.  
These poles are associated with modes that obey outgoing (or radiative) boundary conditions, since as pointed out below (\ref{generalGreens}) they correspond to homogeneous
solutions with the behavior $\sim e^{s(t - \vert x \vert)}$ as $x \rightarrow \pm \infty$.  So quasinormal boundary conditions arise naturally, as a consequence of constructing a retarded Green's function.
The fact that quasinormal boundary conditions arise in this way fits with the intuition that -- barring bound states -- any localized excitation should end up as outgoing radiation.

In this way we are led to construct quasinormal modes with the
peculiar feature that they grow exponentially at spatial infinity.
This follows from the fact that in a stable system the quasinormal
frequencies must have ${\rm Re} \, s < 0$ so that fields decay with
time.  As a result quasinormal modes grow exponentially at spatial
infinity, as follows in general from (\ref{phip}), (\ref{phim}) and
can be seen in the example of a $\delta$-function potential in
(\ref{SingleDeltaQNM}).  Does it make sense to expand a Green's function in terms of modes that grow at spatial infinity?

To see that this isn't a problem, note that causality prevents the exponential growth from showing up in the retarded Green's function.  The retarded Green's function $G(t,x \vert t',x')$
is only non-zero in the future light-cone of $(t',x')$, which
eliminates the problematic regime $\vert x \vert \rightarrow \infty$.
For a single $\delta$-function this can be seen in the last line of (\ref{SingleDeltaG2}),
which indeed grows exponentially with $\vert x \vert$ but gets cut off by causality when $\vert x \vert$ reaches the relevant light cone.  This can also be seen in (\ref{eqn:Gdelta}), where the
step functions cut off the exponential growth with $\vert x \vert$ and in fact require that all of the exponents appearing in (\ref{eqn:Gdelta}) are negative.

Causality plays another important role: over a finite time interval, it limits the way in which the potential can contribute to time evolution.  For example, as pointed out in section \ref{sect:causal}, for two disjoint potentials the sum over echoes is truncated to the maximum number permitted by causality.  A related point is that if the other potential is distant enough to be out of causal contact then it can be ignored.  This is true even though the quasinormal frequencies of the combined system (which are globally-defined quantities that don't know about causality) are very sensitive to distant
perturbations.  
This is relevant to the recent computations of the quasinormal
spectrum of exotic compact objects \cite{Cardoso:2016rao,Cardoso:2016oxy,Ghersi:2019trn,
  Cardoso:2017njb,Wang:2019rcf,
  Price:2017cjr,Mark:2017dnq,Bueno:2017hyj}, where the spectrum differs greatly
from that of the corresponding black hole, yet the initial waveform
(generated by an infalling test particle for example) can be very similar
(e.g. \cite{Cardoso:2016rao}).
The lesson is that over finite times it is not enough
to consider the spectrum of quasinormal frequencies (of the combined
system) by themselves.  One also has to take causality into account.  This can
be done by
making an echo expansion of the form (\ref{eqn:Wexpan}), in which
poles only arise at the quasinormal frequencies of the individual
potentials. 

In this paper we considered potentials with disjoint bounded support to make the analysis tractable.  But causality should be respected in general, with observational as well as theoretical implications.  On the observational side it means that in any practical measurement of the gravitational waves from a merger,
the ringdown will be governed by the quasinormal frequencies of the
resulting black hole and not by distant perturbations to the potential
such as from the surrounding stars (assuming the ringdown can only be
observed for a finite small time interval after merger).
As a theoretical example, for black holes in anti-de Sitter space the Regge-Wheeler potential grows far from
the black hole, and this changes the quasinormal spectrum compared to flat space \cite{Horowitz:1999jd}.  But over short times and distances (much less than an AdS radius) a small black hole in AdS
is not sensitive to the asymptotic potential and to a good approximation perturbations will be governed by the same quasinormal spectrum as in Minkowski space.

We've seen that over a finite time interval distant perturbations to
the potential (meaning perturbations that are out of causal contact)
have no effect.  This leaves the question of the effect of local
perturbations to the potential (meaning perturbations that are
permitted to contribute by causality).  Can we characterize whether local perturbations to the potential have a large or small effect?  This has applications, for example, to observational signatures of
modifications to the near-horizon region of black holes (e.g. \cite{Tattersall:2017erk,Franciolini:2018uyq,McManus:2019ulj}). Although not the main focus of our work, there are some conclusions we can draw.
Consider for example the pair of $\delta$-functions (\ref{DoubleDelta}).  Regarding $V_2$ as a perturbation, and assuming the separation $L$ is small so that $V_2$ is permitted to contribute by causality,
is there a sense in which a small value of $V_2$ has a small effect?  Note that the quasinormal frequencies of the combined system are not a good guide since they change dramatically as soon as $V_2 \not= 0$.
Instead we return to the causal expansion (\ref{eqn:Gdelta}).  Gathering terms with like powers of $V_2$ we see that there are actually two dimensionless expansion parameters.
One dimensionless combination, arising from the exponentials in the numerators, is
\begin{equation}
V_2 \big(t - t_{\rm lc}\big)
\end{equation}
Here $t_{\rm lc}$ is the time at which the relevant light cone first reaches the point $x$.\footnote{In other words, the time at which the argument of the relevant $\theta$ function becomes positive.
For example, for the first sum in (\ref{eqn:Gdelta}), this would be $t = t' + 2 k L + \vert x \vert + \vert x' \vert$.}  Another dimensionless combination, arising from the denominators in (\ref{eqn:Gdelta}),
is $V_2 / V_1$.  So for $V_2$ to have a small effect we require both
\begin{equation}
{V_2 \over V_1} \ll 1 \qquad V_2 \big(t - t_{\rm lc}\big) \ll 1
\end{equation}
The first condition is not surprising; it just says that $V_1$ is the dominant potential.  Assuming that's the case, to interpret the second condition, what value should we take for $t - t_{\rm lc}$?  Since $V_1$ dominates, the
signal which is initially detected decays on a timescale set by the quasinormal frequencies of $V_1$.  For a realistic system (not a $\delta$-function potential) these frequencies are set by the light-crossing time
for the system.\footnote{For example for a black hole they are set by the Schwarzschild radius.}  So in deciding whether the perturbation $V_2$ can appreciably affect the initial signal,
the relevant control parameter (besides $V_2 \ll V_1$) is
\begin{equation}
V_2 \times \hbox{\rm (light-crossing time)}
\end{equation}
If this parameter is small then the initial signal is not appreciably distorted by the perturbation.  Of course even if this parameter is small the perturbation produces late-time
echoes that could in principle be detected.

\bigskip
\goodbreak
\centerline{\bf Acknowledgements}

We thank Emanuele Berti, Vitor Cardoso, Austin Joyce, Alberto Nicolis, Paolo Pani, Rachel Rosen, Luca
Santoni and Michael Zlotnikov for discussions.
LH acknowledges support from NASA NXX16AB27G and DOE DE-SC011941.
DK thanks the Columbia University Center for Theoretical Physics for hospitality during this work.
The work of DK is supported by U.S.\ National Science Foundation grant PHY-1820734.
SW is supported in part by the Croucher Foundation.

\appendix
\section{Green's function solution to the wave equation \label{appendix:Greensolution}}
In this appendix we collect some properties of Green's functions as applied to wave equations.

We first recall how a Green's function can be used to evolve Cauchy data.
Note that if $[\partial_t^2 - \partial_x^2 + V(x)] G(t, x | t', x') =
\delta(t-t') \delta(x-x')$, then $[\partial_{t'}^2 - \partial_{x'}^2 +
V(x')] G(t, x'|t', x) = \delta(t-t') \delta(x-x')$. 
(Note also $G$ is a function of $t-t', x, x'$ alone.)
Applying the second equation to the quantity
\begin{eqnarray}
\int_{t_0}^{\infty} \int_{-\infty}^\infty dx' \phi(t', x')
  [\partial_{t'}^2 - \partial_{x'}^2 + V(x')] G(t, x'|t',x) - 
G(t, x'|t', x) [\partial_{t'}^2 - \partial_{x'}^2 + V(x')]\phi(t', x')
\nonumber \, ,
\end{eqnarray}
with $t_0 < t$, one can see that this quantity is $\phi(t, x)$.
On the other hand, integrating by parts, and assuming
$G$ and $\phi$ vanishes as $x' \rightarrow \pm \infty$, and $G$
and $\partial_{t'} G$ vanishes as $t' \rightarrow \infty$, one sees
that the same quantity gives\footnote{This equality is a statement of Green's formula.}
\begin{eqnarray}
\int_{-\infty}^\infty dx' [G(t, x'|t_0, x) \partial_{t_0} \phi(t_0,
  x') - \phi(t_0, x') \partial_{t_0} G(t, x'|t_0, x)] \, .\nonumber
\end{eqnarray}
The Green's function satisfies reciprocity,\footnote{The proof follows from Green's formula applied to a pair of Green's functions, or alternatively from the observation that the
time-independent Green's function (\ref{1DG}) is symmetric.}
\begin{equation}
G(t,x' \vert t_0,x) = G(t,x \vert t_0,x')
\end{equation}
which lets us switch $x$ and $x'$.  Making this switch, and relabeling $t_0$ as $t'$, yields Eq. (\ref{GreenEvolve}). 

Next we check that Fourier analysis gives rise to an equivalent formula for evolving Cauchy data.  For simplicity we specialize to the free wave equation\footnote{Analogous results in the presence of a potential could be obtained by expanding in eigenfunctions of the relevant Sturm-Liouville operator.}
\begin{eqnarray}
(\partial_t^2 - \partial_x^2 ) \phi(t, x) = 0
\end{eqnarray}
It is easy to see that a given Fourier mode must have either cosine or
sine dependence on time $t$, and thus the evolution from $t'$ to $t$
is given by:
\begin{eqnarray}
\phi(t, x) = \int {dk \over 2\pi} e^{-ikx} 
\left( \phi(t', k) {\,\rm cos\,}(k[t-t']) + \partial_{t'} \phi(t', k)
  \, {{\,\rm sin\,}(k[t-t']) \over k} \right) \, ,
\end{eqnarray}
where $\phi(t', k) = \int dx' \phi(t', x') e^{ik'x'}$. 
To see that this is consistent with the general evolution formula
Eq. (\ref{GreenEvolve}), with
the free retarded Green's function $(1/2) \theta(t - t' - |x - x'|)$,
it is useful to note that
\begin{eqnarray}
\int {dk\over 2\pi} \theta(t-t') {{\,\rm sin\,}(k[t-t']) \over k}
e^{ik(x-x')} = {1\over 2} \theta(t-t'-|x-x'|) \, ,
\end{eqnarray} 
which can be derived by rewriting
$k^{-1} {\,\rm sin\,}(k[t-t']) = \int_{-(t-t')}^{t-t'} dy \,
e^{iky}/2$. This is also consistent with the fact that
$\theta(t-t') {{\,\rm sin\,}(k[t-t']) \over
  k}$ solves $(\partial_t^2 + k^2) G(t-t', k) = \delta(t-t')$.

\section{An alternate method for a multiple $\delta$-function potential\label{appendix:alternate}}
In this appendix we provide the exact solution for the Green's function of (\ref{eqn:wave}) with a multiple $\delta$-function potential, 
\begin{align} \label{eqn:greeneqn}
& \left(\partial_t^2 - \partial_x^2 + \sum_{i}V_i \delta(x-x_i)\right) G(t,x \vert t',x') = \delta(t-t')\delta(x-x'),  \\
& \qquad G(t,x \vert t',x')=   \partial_t G(t,x \vert t',x')=0 \quad  \mbox{for} \quad t<t'
\end{align}
With a Laplace transform in $t - t'$ and Fourier transform in $x$, $\tilde{G}_{s,k}(x') = \int_{0^-}^{\infty} dt \int_{-\infty}^{\infty}dx e^{-st +ikx} G(t,x|0,x') $,  
\begin{align}
 (s^2+k^2)\tilde{G}_{s,k}(x') + \sum_{i} V_i e^{i k x_i} G(s; x_i \vert x') = e^{ikx'}.
\end{align}
After inverse Fourier transform and solving for $G(s; x \vert x')$ we obtain 
\begin{align}
  G(s; x \vert x') =\frac{ e^{-s|x-x'|}}{2s} - \sum_{i}V_i G(s; x_i \vert x') \frac{e^{-s|x-x_i|}}{2s},
\end{align}
where $G(s; x_i \vert x')$ is solved as 
\begin{align}
  G(s; x_i \vert x') = \sum_j (M^{-1})_{ij}  \frac{e^{-s |x_j -x'|}}{2s}, \nonumber \\ 
  M_{ij} = \left\{ \begin{array}{ll}
 1+ \frac{V_i}{2s} & i=j \\   V_j\frac{e^{-s|x_j-x_i|}}{2s} & i\neq j
\end{array}  \right.  .
\end{align}
With a potential composed of two delta functions, $V(x) = V_1 \delta(x) + V_2 \delta(x - L)$, one can easily obtain 
\begin{align}
 G(s; x \vert x') &= \frac{1}{2s}e^{-s|x-x'|}  \nonumber \\
     & \quad + \frac{ V_1(2s+V_2)e^{-s (|x|+ |x'|)} - V_1V_2 e^{-s(|x| + |L|+|L-x'|)} }{2s\left[V_1 V_2 e^{-2s|L|}- (2s +V_1 )(2s+V_2)  \right]}  \nonumber \\
       & \quad + \frac{ V_2(2s+V_1)e^{-s (|L-x|+ |L-x'|)} - V_1V_2 e^{-s(|L-x| + |L|+|x'|)} }{2s\left[V_1 V_2 e^{-2s|L|}- (2s +V_1 )(2s+V_2)  \right]} ,
\end{align}
This reproduces (\ref{eqn:Gs}) without having to solve for $\phi_{+}$, $\phi_{-}$.

\section{WKB approximation for the matching coefficients\label{appendix:ratio}}
In this appendix we use the WKB approximation to obtain an estimate for the matching coefficients (\ref{match}) at large positive ${\rm Re} \, s$ and show that their ratios satisfy (\ref{eqn:ratioasymp}).

For a potential with compact support, ${\rm supp} \, V = (x_1,x_2)$, we want to solve
\begin{equation}
\left(-\partial_x^2 + s^2 + V(x)\right) \phi(x) = 0
\end{equation}
If $s$ is large the effective potential $s^2 + V(x)$ varies adiabatically with $x$.  Then the WKB approximation should be valid, and we can write down a WKB
solution with the behavior (\ref{phip}) needed for $\phi_+$.
\begin{equation}
\phi(x) = \left\lbrace
\begin{array}{ll}
e^{sx} & \quad x < x_1 \\
a \, e^{\int_{x_1}^x dx' \, \sqrt{s^2 + V(x')}} + b \, e^{- \int_{x_1}^x dx' \, \sqrt{s^2 + V(x')}} & \quad x_1 < x < x_2 \\
m^{++} e^{sx} + m^{-+} e^{-sx} & \quad x > x_2
\end{array}\right.
\end{equation}
Requiring that $\phi(x)$ and $\partial_x \phi(x)$ be continuous at $x_1$ and $x_2$ fixes the coefficients, in particular
\begin{eqnarray}
\nonumber
m^{++} & = & {1 \over 4} e^{-s(x_2 - x_1)} e^{\int_{x_1}^{x_2} dx' \, \sqrt{s^2 + V(x')}} \left(1 + {s \over \sqrt{s^2 + V(x_1)}}\right) \left(1 + {\sqrt{s^2 + V(x_2)} \over s}\right) \\
& & \!\!\!\!\!+\, {1 \over 4} e^{-s(x_2 - x_1)} e^{-\int_{x_1}^{x_2} dx' \, \sqrt{s^2 + V(x')}} \left(1 - {s \over \sqrt{s^2 + V(x_1)}}\right) \left(1 - {\sqrt{s^2 + V(x_2)} \over s}\right) \\
\nonumber
m^{-+} & = & {1 \over 4} e^{s(x_1 + x_2)} e^{\int_{x_1}^{x_2} dx' \, \sqrt{s^2 + V(x')}} \left(1 + {s \over \sqrt{s^2 + V(x_1)}}\right) \left(1 - {\sqrt{s^2 + V(x_2)} \over s}\right) \\
& & \!\!\!\!\!+\, {1 \over 4} e^{s(x_1 + x_2)} e^{-\int_{x_1}^{x_2} dx' \, \sqrt{s^2 + V(x')}} \left(1 - {s \over \sqrt{s^2 + V(x_1)}}\right) \left(1 + {\sqrt{s^2 + V(x_2)} \over s}\right)
\end{eqnarray}
As ${\rm Re} \, s \rightarrow + \infty$ this means
\begin{equation}
m^{++} \,\sim\, 1 \hspace{2cm} m^{-+} \,\sim\, e^{2sx_2}
\end{equation}
where we are keeping track of the leading exponential dependence on $s$.  A similar calculation for $\phi_-$ gives
\begin{eqnarray}
\nonumber
m^{+-} & = & {1 \over 4} e^{-s(x_1 + x_2)} e^{\int_{x_1}^{x_2} dx' \, \sqrt{s^2 + V(x')}} \left(1 - {s \over \sqrt{s^2 + V(x_1)}}\right) \left(1 + {\sqrt{s^2 + V(x_2)} \over s}\right) \\
& & \!\!\!\!\!+\, {1 \over 4} e^{-s(x_1 + x_2)} e^{-\int_{x_1}^{x_2} dx' \, \sqrt{s^2 + V(x')}} \left(1 + {s \over \sqrt{s^2 + V(x_1)}}\right) \left(1 - {\sqrt{s^2 + V(x_2)} \over s}\right)
\end{eqnarray}
and implies
\begin{equation}
m^{+-} \,\sim\, e^{-2sx_1}
\end{equation}
Given these results, which can be applied to $V_1$ and $V_2$ separately, the ratios (\ref{eqn:ratioasymp}) follow.

\bibliographystyle{utphys}
\addcontentsline{toc}{section}{References}
\bibliography{QNMechos}

\end{document}